\documentstyle[12pt]{article}
\newcommand{\lpar}{\stackrel{\leftarrow}{\partial}}
\newcommand{\rpar}{\stackrel{\rightarrow}{\partial}}
\begin{document}
\renewcommand{\thefootnote}{\fnsymbol{footnote}}
\begin{center}
{\large\bf
ON ACTION WITH GRASSMANN-ODD LAGRANGIAN\\ }
\vspace{1cm} Vyacheslav A. Soroka
\footnote{E-mail:  vsoroka@kfti.kharkov.ua}
\vspace{1cm}\\
{\it Kharkov Institute of Physics and Technology}\\
{\it 310108, Kharkov, Ukraine}\\
\vspace{1.5cm}
\end{center}
\begin{quotation}
{\small\rm The action with the Grassmann-odd Lagrangian for the
supersymmetric classical Witten mechanics is constructed. It is shown
that the exterior differential can be used for the
connection between Grassmann-even and Grassmann-odd formulations of the
classical dynamics in the superspace in both Hamilton's and Lagrange's
approaches.}
\end{quotation}

\medskip
PACS: 11.15-q, 11.17+y
\renewcommand{\thefootnote}{\arabic{footnote}}
\setcounter{footnote}0

\vspace{1cm}
{\bf 1.\/} There is a well-known close inter-relation between Hamilton's
and Lagrange's formulations of the classical dynamics in the superspace
when both of them are performed by means of the even (with respect to the
Grassmann grading) attributes: the even Poisson bracket and the even
Hamiltonian for the first formulation, and the action with the even
Lagrangian for the last one.  On the other hand, it was shown that
Hamilton's dynamics for the systems, having an equal number of pairs of
even and odd canonical variables, has equivalent Grassmann-odd Hamilton's
formulation with the odd both Poisson bracket and Hamiltonian
\cite{vpst,s}. Therefore, by analogy with the even case it was
naturally assumed \cite{s} that Grassmann-odd Lagrange's formulation of the
dynamics for such systems has also to exist.

In this note the idea about the dynamics formulation based
on the action with the Grassmann-odd Lagrangian is realized on the
example of $d = 1, N = 2$ supersymmetric Witten's mechanics \cite{w}
in its classical version \cite{ap,vpst}.  Note, that the validity of the
statement made in \cite{s} concerning the existence  of the dynamics
formulation by means of  the odd Lagrangian was also confirmed in \cite{f}.

{\bf 2.\/}
Let us consider a system invariant with respect to the $N = 2\
(\alpha = 1,2)$ supersymmetry of the proper time $t$
\par $$
t' = t + i\epsilon^\alpha \theta^\alpha\ , \qquad
\theta'^\alpha\ = \theta^\alpha + \epsilon^\alpha\ .
$$
By using the covariant derivatives
\par $$
D_\alpha = {\partial \over \partial \theta^\alpha} - i
\delta_{\alpha\beta}\theta^\beta {\partial \over \partial t}\ ,
$$
and two real scalar superfields
\par $$
\Phi(t,\theta^1,\theta^2) = q(t) +
i {\psi_\alpha}(t) \theta^\alpha + i F(t) \theta^\alpha\theta_\alpha \ ,
$$
$$
\Psi(t,\theta^1,\theta^2) = \eta(t) + i {a_\alpha}(t) \theta^\alpha
 + i \Xi(t) \theta^\alpha\theta_\alpha       \ ,
$$
having the opposite values of the Grassmann grading $g$ $(g(\Phi) = 0,
g(\Psi) = 1)$, the following supersymmetric action $\bar S$ with the
Grassmann-odd Lagrangian $\bar L$ $(g(\bar L) = 1)$ can be constructed
\par $$
\bar S = \int dtd\theta_2d\theta_1
\left[ - {1 \over2} D^\alpha \Psi D_ \alpha \Phi + i \Psi W(\Phi)\right] =
\int dt \bar L\ ,
\eqno {(1)} $$
where $W(\Phi)$ is an arbitrary real function
of $\Phi$ and the index $\alpha$ is raised $D^\alpha =
\epsilon^{\alpha\beta}D_\beta$ and lowered $\theta_\alpha =
\epsilon_{\alpha\beta}\theta^\beta$ by means of the antisymmetric tensors
$\epsilon^{\alpha\beta}$, $\epsilon_{\alpha\beta}$
$(\epsilon^{\alpha\beta}\epsilon_{\beta\gamma} = \delta^\alpha_\gamma,\
\epsilon^{12} = 1)$.
Excluding after integration in (1) over the Grassmann variables
$\theta^1, \theta^2$ the auxiliary fields $F$ and $\Xi$
with the help of their equations of motion, we obtain
\par$$
\bar L =\ \stackrel{\cdot}{\eta} \stackrel{\cdot}{q} +
{i \over 2} ( a^\alpha\stackrel{\cdot}{\psi^\alpha} -
\stackrel{\cdot}{a^\alpha}\psi^\alpha - 2 a^\alpha \psi_\alpha W') -
\eta\ ( WW' + {i \over 2} \psi^\alpha \psi_\alpha W'')\ ,
\eqno{(2)} $$
where the dot and the prime mean the derivatives with respect to
$t$ and $q$ correspondingly. The odd Lagrangian $\bar L$ leads to
the momenta
\par$$
p = {\partial \bar L \over \partial \stackrel{\cdot}{\eta}}
=\ \stackrel{\cdot}{q}\ ; \qquad\pi = {\partial \bar L \over \partial
\stackrel{\cdot}{q}} =\ \stackrel{\cdot}{\eta}\ ;
\eqno{(3a)} $$
\par$$
\pi^\alpha = {\partial \bar L \over
\partial\stackrel{\cdot}{a^\alpha}} =\ -{i \over 2} \psi^\alpha\ ;\qquad
p^\alpha = {\partial \bar L \over \partial\stackrel{\cdot}{\psi^\alpha}}
=\  {i \over 2} a^\alpha\ ,
\eqno{(3b)} $$
canonically conjugate to the coordinates $\eta$, $q$, $a^\alpha$
and $\psi^\alpha$ in the odd bracket
\par $$
\{\eta, p\}_1 = \{q, \pi \}_1 = 1; \qquad
\{a^\alpha, \pi^\beta \}_1 = \{\psi^\alpha, p^\beta \}_1 =
\delta^{\alpha \beta}\ ,
\eqno{(4)} $$
where the remaining odd-bracket relations between the canonical variables
with zero in their right-hand sides are unwritten here.  Relations (3a)
express the velocities $\stackrel{\cdot}{q}$ and $\stackrel{\cdot}{\eta}$
in terms of the corresponding momenta, while (3b) define the constraints
\par $$
\varphi^\alpha = \pi^\alpha + {i
\over 2} \psi^\alpha\ ;\qquad f^\alpha = - p^\alpha + {i \over 2}
a^\alpha\ ,
\eqno{(5)} $$
which, as can be verified by using
relations (4), are of the second class
\par $$
\{ \varphi^\alpha, f^\beta
\}_1 = - i \delta^{\alpha \beta} ;\qquad \{ \varphi^\alpha, \varphi^\beta
\}_1 = \{ f^\alpha, f^\beta \}_1 = 0\ .
\eqno{(6)} $$
We designate the Grassmann-even coordinates, momenta and constraints with
the Latin letters, while the odd ones with the Greek letters. If we
introduce the variables
\par$$
\chi^\alpha = \pi^\alpha - {i \over 2}
\psi^\alpha\ ;\qquad g^\alpha = - p^\alpha - {i \over 2} a^\alpha\ ,
$$
which are canonically conjugate each other in the odd bracket (4)
\par$$
\{ \chi^\alpha, g^\beta \}_1 = i
\delta^{\alpha \beta} ;\qquad
\{ \chi^\alpha, \chi^\beta \}_1 =
\{ g^\alpha, g^\beta \}_1 = 0\ ,
$$
and whose odd-bracket relations with the constraints $\varphi^\alpha$,
$f^\alpha$ are vanished, then Dirac's odd bracket from any
functions $A$ and $B$ take the form
\par$$
\{A, B\}_1^{D.B.} =
\{A, B\}_1 + i \{A, \varphi^\alpha\}_1\{f^\alpha, B\}_1 -
i \{A, f^\alpha\}_1\{\varphi^\alpha, B\}_1 =
$$
$$
= A \left({\lpar}_q{\rpar}_\pi - {\lpar}_\pi{\rpar}_q +
{\lpar}_\eta{\rpar}_p - {\lpar}_p{\rpar}_\eta +
i {\lpar}_{\chi^\alpha}{\rpar}_{g^\alpha} -
i {\lpar}_{g^\alpha}{\rpar}_{\chi^\alpha} \right) B\ ,
\eqno{(7)} $$
where ${\lpar}$ and
${\rpar}$ are  right and left derivatives, and
the notation $\partial_z = {\partial \over \partial z}$ is introduced.
Following from the Lagrangian (2) the total odd Hamiltonian,
if subjected to the second-class constraints $\varphi^\alpha = 0$
and $f^\alpha = 0$, takes the form
\par$$
\bar H = p\pi + \eta\ (WW' + {i \over 2} \chi_\alpha
\chi^\alpha W'') + i g_\alpha\chi^\alpha W'
\eqno{(8)} $$
and with the use of Dirac's bracket (7) gives Hamilton's equations
\par$$
\stackrel{\cdot}{x^a} = \{x^a, \bar H \}_1^{D.B.}
$$
for the independent phase variables $x^a = (q, p, \chi^\alpha,
\eta, \pi, g^\alpha)$
\par$$
\stackrel{\cdot}{q} = p\ ,\qquad
\stackrel{\cdot}{p} = -\ WW' - {i \over 2} \chi_\alpha \chi^\alpha W''\
,\qquad \stackrel{\cdot}{\chi^\alpha} = \chi_\alpha W'\ ,
\eqno{(9a)} $$
$$
\stackrel{\cdot}{\eta} = \pi\ ,\qquad
\stackrel{\cdot}{\pi} = -\ \left\{\ {\eta \over 2}\ [(W^2)'' +
\chi_\alpha \chi^\alpha W'''] + i g_\alpha \chi^\alpha W''\right\}\ ,\qquad
\stackrel{\cdot}{g^\alpha} = g_\alpha W' + \eta \chi_\alpha W''\ .
\eqno{(9b)} $$
Equations (9a) are Hamilton's equations for Witten's
supersymmetric mechanics \cite{w} in its classical version \cite{vpst}
which can be derived by means of Dirac's even bracket
\par
$$
\{A, B\}_0^{D.B.} =
\{A, B\}_0 - i \{A, \varphi^\alpha\}_0\{\varphi^\alpha, B\}_0 =
A \left({\lpar}_q{\rpar}_p - {\lpar}_p{\rpar}_q +
i {\lpar}_{\chi^\alpha}{\rpar}_{\chi^\alpha} \right) B
\eqno{(10)} $$
with the help of the even Hamiltonian $H$
\par $$
H = {{p^2 + W^2(q)}\over 2} + {i \over 2} \chi_\alpha \chi^\alpha W'(q)\ ,
\eqno{(11)} $$
which both follow from the $N = 2$ supersymmetric action
with the Grassmann-even Lagrangian L $(g(L) = 0)$ (see, for example,
\cite{ap})
\par $$
S = {1 \over 4} \int dtd\theta_2d\theta_1
\left[ D^\alpha \Phi D_ \alpha \Phi + 2i V(\Phi)\right] =
\int dt  L\ ,
\eqno {(12)} $$
where $V'(\Phi) = 2W(\Phi)$ and the even Lagrangian after exclusion of the
auxiliary field $F$ is
\par $$
L = {1\over 2}\ [ \stackrel{\cdot}{q}^2
+\ i ( \psi^\alpha\stackrel{\cdot}{\psi^\alpha}
+\ \psi_\alpha \psi^\alpha W') - W^2 ]\ .
\eqno {(13)} $$
The momenta canonically conjugate to the coordinates $q$ and
$\psi^\alpha$ in the even bracket
\par$$
\{q, p\}_0 = 1; \qquad \{\psi^\alpha, \pi^\beta \}_0 =
- \delta^{\alpha \beta}\ ,
$$
following from the even Lagrangian (13), have the form
\par$$
p = {\partial L \over \partial \stackrel{\cdot}{q}}
=\ \stackrel{\cdot}{q}\ ; \qquad
\pi^\alpha = {\partial L \over
\partial\stackrel{\cdot}{\psi^\alpha}} =\ -{i \over 2} \psi^\alpha\ .
$$
The last relations define the second-class constraints
\par$$
\varphi^\alpha = \pi^\alpha + {i \over 2} \psi^\alpha\ ;\qquad
\{ \varphi^\alpha, \varphi^\beta \}_0 = - i
\delta^{\alpha \beta}\ ,
\eqno {(14)} $$
which commute in the even bracket with the variables
\par$$
\chi^\alpha = \pi^\alpha - {i \over 2} \psi^\alpha\ ;\qquad
\{ \chi^\alpha, \chi^\beta \}_0 = i
\delta^{\alpha \beta}\ ,
$$
entering into the definitions for the even Dirac bracket (10) and the even
Hamiltonian (11). The even Hamiltonian (11) follows with the use of the
second-class constraints restriction $\varphi^\alpha = 0$ from
the total even Hamiltonian corresponding to the Lagrangian (13).

Equation (9b) can be obtained by taking
the exterior differential $d$ of the Hamilton equations (9a) for the
Witten mechanics and performing the map $\lambda$:
\par $$
dq \rightarrow \eta ;\qquad dp \rightarrow \pi ;\qquad
d\chi^\alpha \rightarrow g^\alpha
$$
$$
d\psi^\alpha \rightarrow a^\alpha ;\qquad
d\pi^\alpha \rightarrow - p^\alpha ;\qquad
dF \rightarrow \Xi ;\qquad
d\varphi^\alpha \rightarrow f^\alpha\ .
\eqno{(15)} $$
Note, that we identify the grading of the exterior differential
$d$ with the Grassmann grading $g$ of the quantities
$\theta^\alpha$ ($g(d) = g(\theta^\alpha) = 1$), i.e., $g(dx^a) = g(x^a) +
1$.
The composition $\lambda \circ d$ of the maps $\lambda$ and $d$  renders
the even Hamiltonian (11) into the odd one (8)
\par $$
dH \stackrel{\lambda}{\rightarrow} \bar H $$

{\bf 3.\/}
The inter-relation of the brackets (10), (7) and of the corresponding to
them Hamiltonians (11), (8) can be described by the
following scheme. Hamilton's equations of a system
expressed by means of the usual even Poisson-Martin bracket with
the use of the Grassmann-even Hamiltonian $H$
\par$$
\stackrel{\cdot}{x^a} = \{x^a, H \}_0 =\
 \omega^{ab} {{\partial {H}} \over \partial{x^b}} \ ,
\eqno{(16)} $$
can be rewritten as
\par$$
\stackrel{\cdot}{x^a} =\
 \omega^{ab} {{\partial {H}} \over \partial{x^b}} \equiv\
 \omega^{ab} {{\partial {(dH)}} \over \partial{(dx^b)}}
  \stackrel{\rm def}{=} \{x^a, dH \}_1\ .
\eqno{(17)} $$
The exterior differential of the Hamilton equations (16) can be expressed
in the form
\par$$
\stackrel{\cdot}{dx^a} = (d\omega^{ab}) {{\partial {(dH)}}
\over \partial{(dx^b)}} + (-1)^{g(a) + g(b)} \omega^{ab}
d({{\partial {H}} \over \partial{x^b}})\ .
\eqno{(18)} $$
If $x^a$ and $dx^a$ are considered as independent variables, then
the order of the exterior and partial differentiations can be changed in
the second term in the right-hand side of relation (18) which, thus,
take the form
\par$$
\stackrel{\cdot}{dx^a} = (d\omega^{ab}) {{\partial {(dH)}}
\over \partial{(dx^b)}} + (-1)^{g(a) } \omega^{ab}
{{\partial {(dH)}} \over \partial{x^b}} \stackrel{\rm def}{=}
\{dx^a, dH \}_1\ .
$$
Thus, we introduced by definition the odd bracket
\par $$
\{ A , B \}_1 =
$$
$$
= A \left[
\stackrel{\leftarrow}{{\partial}\over \partial{x^a}}
\omega^{ab}
\stackrel{\rightarrow}{{\partial} \over \partial{(dx^b)}} +
(-1)^{g(a)}
\stackrel{\leftarrow}{{\partial} \over \partial{(dx^a)}}
\omega^{ab}
\stackrel{\rightarrow}{{\partial} \over \partial{x^b}} +
\stackrel{\leftarrow}{{\partial}\over \partial{(dx^a)}}
(d{\omega}^{ab})
\stackrel{\rightarrow}{{\partial} \over \partial{(dx^b)}}\right] B\ ,
\eqno {(19)} $$
that reproduces with the use of the odd Hamiltonian $dH$ the Hamilton
equations (17) for the phase coordinates $x^a$, obtained by means of the
even bracket with the help of the even Hamiltonian $H$, and gives,
besides, the equations for $dx^a$. Note, that the last term in the
definition (19) disappears if we use a canonical form for the
even bracket (16). As can be verified with the use of the
properties of the matrix $\omega^{ab}$ entering in the even bracket (16),
the bracket (19) does possess the properties necessary for the odd
bracket. In connection with a similar scheme see also the paper \cite{n}.

It is interesting to note that, by using this scheme, the Grassmann-odd
Hamilton formulation for the supersymmetric one-dimensional oscillator
fulfilled by means of the odd bracket
\par$$
\stackrel{\cdot}{x^a} = \{x^a, \bar H \}_1 =
 x^a \left({\lpar}_{\theta^1}{\rpar}_{q} - {\lpar}_{q}{\rpar}_{\theta^1} +
{\lpar}_{\theta^2}{\rpar}_{p} - {\lpar}_{p}{\rpar}_{\theta^2} \right) \bar H
$$
with the odd Hamiltonian $\bar H = q \theta^2 - p \theta^1$
can be obtained from the even Hamilton formulations of both
the Bose-oscillator and the Fermi-oscillator which are described
correspondingly in the Poisson and Martin even brackets
\par$$
\{A, B\}_0^{P.B.} =
A \left( {\lpar}_{q} {\rpar}_{p} - {\lpar}_{p} {\rpar}_{p}  \right) B ;
\qquad
\{A, B\}_0^{M.B.} =
- i  A \left({\lpar}_{\theta^1}{\rpar}_{\theta^1} +
{\lpar}_{\theta^2}{\rpar}_{\theta^2} \right) B
$$
with the even Hamiltonians $ H = (p^2 + q^2) / 2$ and
$\tilde H = i \theta^1 \theta^2$, respectively.
Under this, the operation of taking the exterior differential must be
accompanied by the designations $dq = \theta^2$, $dp = -\ \theta^1$ in the
Bose-oscillator case and $i d\theta^1 = q$, $i d\theta^2 = p$ for the
Fermi-oscillator.

In general case if we have a Hamilton system which dynamics
is described by means of the bracket $\{ A , B \}_{\epsilon }$
with the help of the Hamiltonian $\stackrel{\epsilon}{H}$
\par$$
\stackrel{\cdot}{x^a} = \{x^a, \stackrel{\epsilon}{H} \}_{\epsilon } =\
 \stackrel{\epsilon}{\omega}^{ab} {{\partial {\stackrel{\epsilon}{H}}}
\over \partial{x^b}} \ ,
\eqno{(20)} $$
where $\epsilon$ $(\epsilon = 0,1)$
is the Grassmann parity of both the bracket and the Hamiltonian, then the
Hamilton equations for the phase coordinates $x^a$ and the equations for
their differentials $dx^a$, obtaining by a differentiation of equations
(20), can be reproduced by the following bracket of the opposite
Grassmann parity
\par $$
\{ A , B \}_{\epsilon + 1} =
$$ $$
= A \left[
\stackrel{\leftarrow}{{\partial}\over \partial{x^a} }
\stackrel{\epsilon}{\omega}^{ab}
\stackrel{\rightarrow}{{\partial} \over \partial{(dx^b)}} +
(-1)^{g(a) + \epsilon}
\stackrel{\leftarrow}{{\partial} \over \partial{(dx^a)}}
\stackrel{\epsilon}{\omega}^{ab}
\stackrel{\rightarrow}{{\partial} \over \partial{x^b}} +
\stackrel{\leftarrow}{{\partial}\over \partial{(dx^a)} }
(d\stackrel{\epsilon}{\omega}^{ab})
\stackrel{\rightarrow}{{\partial} \over \partial{(dx^b)}}\right] B
\eqno {(21)} $$
with the help of the Hamiltonian $d{\stackrel{\epsilon}{H}}$
$(g(d{\stackrel{\epsilon}{H}}) = \epsilon + 1)$, that is,
\par$$
\stackrel{\cdot}{x^a} = \{x^a, \stackrel{\epsilon}{H} \}_{\epsilon } =\
 \{x^a, d{\stackrel{\epsilon}{H}} \}_{\epsilon + 1}\ ;\qquad
\stackrel{\cdot}{dx^a} = d(\{x^a, \stackrel{\epsilon}{H} \}_{\epsilon }) =\
\{dx^a, d{\stackrel{\epsilon}{H}} \}_{\epsilon + 1}\ .
$$
We can again verify, by using the properties of
the matrix $\stackrel{\epsilon}{\omega}^{ab}$ for the bracket (20), that
(21), in fact, satisfies all the properties necessary for the bracket
$\{ A , B \}_{\epsilon + 1}$.

As in the case of Hamilton's formulations there is an
interconnection between Lagrange's equations corresponding to
the Lagrangians of the different Grassmann parities $L$ (13) and $\bar L$
(2), because the odd Lagrangian $\bar L$ (2) is related by means of the
redefinition $\lambda$ (15) with the exterior differential $dL$ of the
even Lagrangian (13).  Indeed, Lagrange's equations corresponding to the
Lagrangian $\stackrel{\epsilon}{L}(q^a, \stackrel{\cdot}{q^a})$ with the
Grassmann parity $\epsilon$ can be written in the two equivalent forms
\par$$
{d \over dt} \left({\partial{\stackrel{\epsilon}{L}} \over
\partial{\stackrel{\cdot}{q^a}}}\right)
- {\partial{\stackrel{\epsilon}{L}} \over \partial{q^a}} = 0
\Longleftrightarrow
{d \over dt} \left({\partial{(d{\stackrel{\epsilon}{L}})} \over
\partial{(\stackrel{\cdot}{dq^a}})}\right) -
{\partial{(d{\stackrel{\epsilon}{L}})} \over \partial{(dq^a)}} = 0\ ,
\eqno{(22a,b)}
$$
while the equations obtained by taking the differential of (22a) have the
form
\par$$
{d \over dt} \left({\partial{(d{\stackrel{\epsilon}{L}})} \over
\partial{\stackrel{\cdot}{q^a}}}\right) -
{\partial{(d{\stackrel{\epsilon}{L}})} \over \partial{q^a}} = 0\ .
\eqno{(22c)}
$$
Equations (22b) together with (22c) can be considered as Lagrange's
equations for the system described in the configuration space with
the coordinates $q^a$, $dq^a$ by the Lagrangian
$d{\stackrel{\epsilon}{L}}$ of the Grassmann parity $\epsilon + 1$.

Note also that if the Lagrangian $\stackrel{\epsilon}{L}$ has the
constraints $\varphi^i(q^a, p_a = {\partial{\stackrel{\epsilon}{L}}}
 / {\partial{\stackrel{\cdot}{q^a}}})$ satisfying the relations in the
bracket corresponding to $\stackrel{\epsilon}{L}$
\par$$
\{ \varphi^i, \varphi^k \}_{\epsilon} = f^{ik}\ ,
$$
then the Lagrangian $d{\stackrel{\epsilon}{L}}$ will possess the
constraints $\varphi^i(q^a, p_a = {\partial(d{\stackrel{\epsilon}{L}})}
 /  {\partial({\stackrel{\cdot}{dq^a}})})$, coinciding with those
following from $\stackrel{\epsilon}{L}$, and $d\varphi^i$ obeying the
relations
\par$$
\{ \varphi^i, d\varphi^k \}_{\epsilon + 1} = f^{ik};\qquad
\{ \varphi^i, \varphi^k \}_{\epsilon + 1} = 0\ ,
$$
which follow from the related with $d{\stackrel{\epsilon}{L}}$ bracket
expression (21) that in the case is without the last term in their
right-hand side (cf., e.g., equations (14) with (5), (6)).

{\bf 4.\/}
Thus, it is shown that for the given formulation of the dynamics (either
in Hamilton's or in Lagrange's approach) with  the equations
of motion for the dynamical variables $z^a$ we can construct,
by using the exterior differential, such a formulation, having the
opposite Grassmann parity, that reproduces the former equations for $z^a$
and gives, besides, the equations for their differentials $dz^a$.

\bigskip

The author is sincerely thankful to D.V. Volkov for useful discussions.

\medskip
This work was supported in part by the Ukrainian State Committee in
Science and Technologies, Grant N 2.3/664, by Grant N UA 6000 from
the International Science Foundation, by Grant N UA 6200 from Joint
Fund of the Government of Ukraine and International Science Foundation
and by Grant N 93-127 from INTAS.

\end{document}